\documentclass[aps,reprint,prl,superscriptaddress,showpacs]{revtex4-1}

\usepackage{graphicx}
\usepackage{amsmath,graphicx,dcolumn}
\usepackage{hyperref}
\usepackage[usenames]{color}
\usepackage{datetime}

\newcommand*{\etal}{{\em et al.}}
\newcommand*{\ur}{URu$_2$Si$_2$}

\newcommand*{\an}{{${\AA}^{-1}$}}

\begin{document}

\title{Anomalous Femtosecond Quasiparticle Dynamics of Hidden Order State in {\ur} \bigskip}

\author{Georgi L. Dakovski}
\affiliation{Center for Integrated Nanotechnologies, Los Alamos
National Laboratory, Los Alamos, NM 87545, USA}

\author{Yinwan Li}
\affiliation{Condensed Matter and Magnet Science Group, Los Alamos
National Laboratory, Los Alamos, NM 87545, USA}

\author{Steve M. Gilbertson}
\affiliation{Center for Integrated Nanotechnologies, Los Alamos
National Laboratory, Los Alamos, NM 87545, USA}

\author{George Rodriguez}
\affiliation{Center for Integrated Nanotechnologies, Los Alamos
National Laboratory, Los Alamos, NM 87545, USA}

\author{Alexander V. Balatsky}
\affiliation{Theoretical Condensed Matter Physics Group and Center for Integrated Nanotechnologies, Los Alamos
National Laboratory, Los Alamos, NM 87545, USA}

\author{Jian-Xin Zhu}
\affiliation{Physics of Condensed Matter and Complex Systems Group, Los Alamos
National Laboratory, Los Alamos, NM 87545, USA}

\author{Krzysztof Gofryk}
\affiliation{Condensed Matter and Magnet Science Group, Los Alamos
National Laboratory, Los Alamos, NM 87545, USA}

\author{Eric D. Bauer}
\affiliation{Condensed Matter and Magnet Science Group, Los Alamos
National Laboratory, Los Alamos, NM 87545, USA}

\author{Paul H. Tobash}
\affiliation{Condensed Matter and Magnet Science Group, Los Alamos
National Laboratory, Los Alamos, NM 87545, USA}

\author{Antoinette Taylor}
\affiliation{Materials Physics and Applications Division, Los Alamos
National Laboratory, Los Alamos, NM 87545, USA}

\author{John L. Sarrao}
\affiliation{Condensed Matter and Magnet Science Group, Los Alamos
National Laboratory, Los Alamos, NM 87545, USA}

\author{Peter M. Oppeneer}
\affiliation{Department of Physics and Astronomy, Uppsala University, Box 516, S-75120 Uppsala, Sweden }

\author{Peter S. Riseborough}
\affiliation{Department of Physics, Temple University - Philadelphia, PA 19122, USA}

\author{John A. Mydosh}
\affiliation{Kamerlingh Onnes Laboratory, Leiden University, NL-2300 RA Leiden, The Netherlands }

\author{Tomasz Durakiewicz* \bigskip}
\affiliation{Condensed Matter and Magnet Science Group, Los Alamos
National Laboratory, Los Alamos, NM 87545, USA}

\begin{abstract}
At T$_0$ = 17.5 K an exotic phase emerges from a heavy fermion state in {\ur}. 
The nature of this hidden order (HO) phase has so far evaded explanation. 
Formation of an unknown quasiparticle (QP) structure is believed to be responsible for the massive removal of entropy at HO transition, however, experiments and ab-initio calculations have been unable to reveal the essential character of the QP.
Here we use femtosecond pump-probe time- and angle-resolved photoemission spectroscopy (tr-ARPES) to elucidate the ultrafast dynamics of the QP.  
We show how the Fermi surface is renormalized by shifting states away from the Fermi level at specific locations, characterized by vector $q_{<110>} = 0.56 \pm 0.08$ {\an}. 
Measurements of the temperature-time response reveal that upon entering the HO the QP lifetime in those locations increases from 42 fs to few hundred fs. 
The formation of the long-lived QPs is identified here as a principal actor of the HO.
\end{abstract}

\maketitle

Over the last 25 years, the nature of the hidden order transition in a heavy fermion system {\ur} has remained a mystery. The sharp second-order transition at T$_0$= 17.5 K \cite{Palstra,Schlabitz,Maple, Chandra,Santini,Kasuya,Varma,Elgazzar,Harima,Haule,Balatsky,Dubi, Schmidt}, marked by a large jump in specific heat, corresponds to removal of more than 10 percent of
total entropy \cite{Wiebe}. The phase transition might have been consistent with magnetism, but several years of intensive search showed the absence of magnetism in HO phase. Antiferromagnetism is observed only under pressure \cite{Hassinger}. Partial gapping of the Fermi surface was proposed in the context of itinerant models
\cite{Varma,Elgazzar, Balatsky,Dubi} and the HO gap was shown to behave similarly to a BCS order parameter \cite{Aynajian}. Entropy removal at HO transition also
suggested that the Fermi surface instability induces reconstruction of the density of states. It was proposed that a commensurate \cite{Tripathi,Elgazzar,Kasuya,Ikeda,Harima} or incommensurate \cite{Chandra, Balatsky,Dubi} renormalization of the Fermi surface is the driver of HO transition, with magnetic fluctuations\cite{Elgazzar} or hybridization wave \cite{Balatsky,Dubi} as mechanisms for the gap formation. However, the underlying
physics behind the key QPs, the gap formation, symmetry, and momentum-dependence remain elusive, in spite of several models of HO proposed over the years \cite{Chandra,Tripathi,Varma,Elgazzar,Ikeda,Harima,Haule,Dubi}.

\begin{figure}[h]
\includegraphics[width=8.5cm]{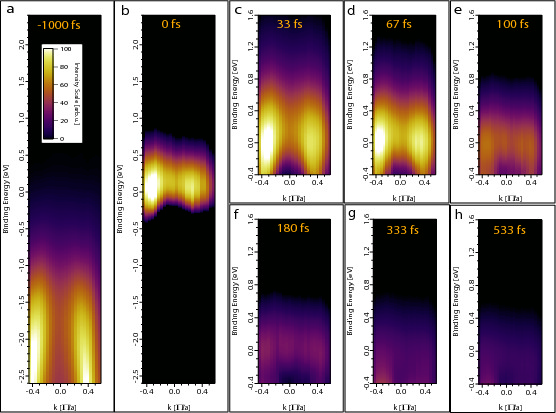}
\caption{(color online) Time-resolved ARPES study of {\ur} in the Hidden Order state. { \bf {a,}} the case of very long, "negative" 1ps delay between probe and subsequent pump pulses. This spectrum is used to subtract a background from all other spectra collected for probe pulse arriving after pump pulse with delays between 0 fs and 533 fs, with { \bf {b,}} "0 fs" case corresponding to the overlap of pump and probe pulses. The double structure seen above the Fermi level corresponds to the new, long-lived QPs, decaying in density between 0 fs and 533 fs, as shown in { \bf {b}} to { \bf {h}}. The characteristic lifetime estimated from the decay in intensity as function of delay is 213 $\pm$ 12fs, corresponding to a QP peak width of 3meV. Long-lived QPs s are located right above the Fermi level and separated by vector $q = 0.56 \pm 0.08$ $\Pi/a$.  All pump-probed angle-resolved spectra shown in this Figure are taken below T$_0$ with variable pump-probe delay.}
\end{figure}

The picture of HO is also obscured by the formation of a hybridization gap, a typical feature of heavy fermion materials \cite{BauerPRL}. Such a gap formation is due to hybridization of the flat f-band with a strongly dispersive d-band, as evidenced by numerous experiments \cite{Santander, Broholm,Wiebe, Palstra,Schlabitz,Maple}. The hybridization gap onset is related to the coherence temperature T*, usually from 60 K to 100 K for uranium-based heavy fermion materials \cite{Yang2} and estimated at 70 K in {\ur}. At T$_0$ the stage is already set by a well established f-d hybridization gap structure evolving in a mean-field behavior \cite{Riseborough, Rise2}.
\begin{figure}
\includegraphics[width=8.5cm]{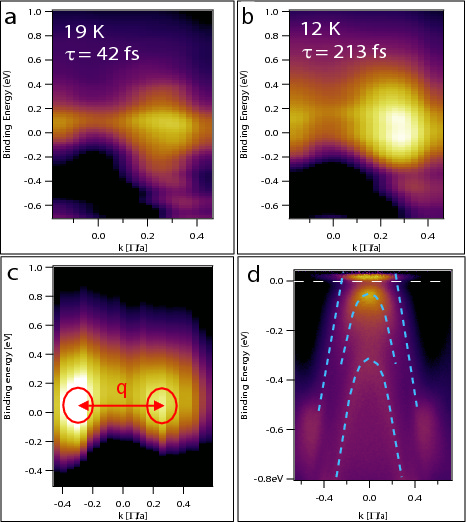}
\caption{(color online) ARPES and tr-ARPES of \ur.  Time-resolved ARPES below and above T$_0$ is shown in panels { \bf {a}} and { \bf {b,}} respectively, measured around zero delay between pump and probe and normalized. We note the increase in QP density and lifetime upon HO transition.{ \bf {c,}} the two QP spots as described in Fig. 1b are shown here, with the location of the vector $q_{<110>} = 0.56 \pm 0.08$ {\an}. For comparison with static photoemission, ARPES scan is shown in { \bf {d,}} with measured intensity divided by Fermi Function. This result provides evidence for the flat band above the Fermi level. ARPES data were collected with 34eV photons at a position in momentum space of the bct Brillouin zone at the Z point with sample temperature of 12K. Dashed lines mark the location of dominant bands in the spectrum. Color intensity scale is the same as in Fig. 1. }
\end{figure}
To reveal the nature of the HO transition within the hybridization gap, one needs to track the QPs responsible for the Fermi surface renormalization. Several recent investigations using ARPES, INS and Scanning Tunneling Microscopy (STM) demonstrate that the HO transition is marked by modification of the density of states \cite{Wiebe,Santander,Schmidt,Aynajian}. The two recent ARPES experiments \cite{Santander, Yoshida} and this work can be reconciled with a three dimensional nature of the electronic structure in {\ur} \cite{adinf}. These findings set the stage for resolving the critical unknowns: what is the nature of the QPs driving the HO transition?  How does the HO gap structure evolve in energy-momentum space? Where does the missing entropy\cite{Wiebe, Tripathi} go? And finally: is the hidden order gap related to the hybridization gap?
\begin{figure}
\includegraphics[width=8.5cm]{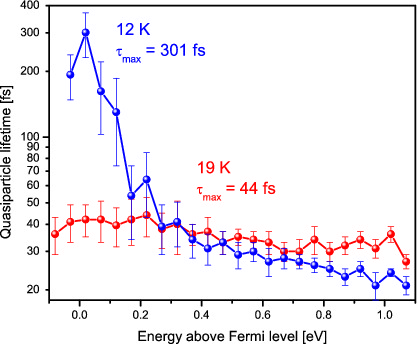}
\caption{
{(color online) Quasiparticle lifetime. The maximum QP lifetime in and above the HO state is shown as a function of energy above the Fermi level. The maximum lifetime situated at the Fermi level corresponds to the formation of the long-lived QPs. These QPs are located at the HO "hotspots", as shown in Figures 1 and 2, and can be identified as the
principal driver and a fingerprint of the HO. We find 301 fs as the maximum QP lifetime at 12 K. This anomalous lifetime corresponds to a QP peak width of 2meV and reflects the HO gap structure. The short-lived  hybridization gap structure is seen above T$_0$, where the lifetime drops to a maximum value of 44fs, with a corresponding width of 15meV. See Supporting Material for more details.}
}
\end{figure}
To answer these questions, the temperature and time structure of the transition-driving QPs needs to be examined. Here we report the novel approach of time-resolved ARPES measurements on the femtosecond scale, measuring the ultrafast QP dynamics in normal and HO states with 29.5 eV probe photons, at a position in momentum space of the bct Brillouin zone along k$_z$ slightly below the Z point. Time-resolved ARPES allows us to populate the empty states with a pump pulse of 1.55 eV and measure the energy and momentum resolved electronic structure of the bands above the Fermi level. The populated electronic structure decays at a rate proportional to QP lifetime limited by electron-electron or electron-phonon scattering processes and its evolution is measured in small time intervals by a probe pulse to trace the decay. Temporal resolution better than 50 fs is needed for analyzing electron-electron processes, and photon energy in the range of cross sections favorable for 5f photoionization is crucial. The tr-ARPES instrument utilizing the high-harmonic generation (HHG) and time-compensating monochromator was constructed specifically for work with f-electron systems, and utilizes the HHG based on noble gas excited by a 10 kHz, 1.55 eV laser with 10 kHz repetition rate, coupled to the time-compensated double monochromator \cite{RSI}. The differential pumping system maintains UHV conditions in the measurement chamber, where photoelectron spectra are acquired with a hemispherical electron energy analyzer.  We use 1.55 eV, 30 fs pump pulses to avoid heating of the electron gas, and 29.5 eV, 15 fs probe pulses to achieve favourable cross section for f-orbital photoionization. ARPES experiments were performed in the Synchrotron Radiation Center, utilizing the PGM beamline 71A and the SCES4000 hemispherical electron energy analyzer, with energy resolution of 15 meV at 34 eV photons. Single crystals of {\ur} were grown via the Czochralski technique in electrical tri-arc and tetra-arc furnaces, followed by a 900 degrees C anneal in Ar for one week in the presence of a Zr getter.

Fig. 1 shows the electronic structure of {\ur} in the $<$110$>$ direction measured with tr-ARPES in the HO state at 12K. Panel a) shows the structure as seen
with the probe pulse arriving 1 ps before the pump pulse. This is the semi-static structure, since the 10kHz repetition rate allows all the states populated
with the previous pump pulse to decay long before the probe pulse arrives. This structure is used as a baseline, and is subtracted from data taken at
other delays, shown in all consecutive panels in Fig. 1. This subtraction allows us to see only the part of electronic structure being populated by
the pump pulse. At delay zero (panel b), one can already see the very well-formed structure at the Fermi level, composed of two QP spots located
around  $k_{HO} = \pm 0.28 \pm 0.04$ {\an} and separated by a vector $q_{<110>}  = 2k_{HO} =  0.56 \pm  0.08$ {\an}. This value agrees with the STM estimate of $k_{HO}=0.3$ {\an} for the k$_z$-integrated case. In time, as seen in panels c) to h), the QP density of states decreases, with the characteristic time constant of 213 $\pm$ 13 fs. These long-lived QPs are located around well-defined points at the upper edge of the HO gap, corresponding to the areas of highest susceptibility, where formation of f-electron QPs drives the HO. The appearance of such localized, long-lived QPs is consistent with the partial gapping and removal of DOS from the Fermi level, and consequently, entropy, at T$_0$. Observation of QPs localized in momentum space and with a dramatically increased lifetime is the central result of this paper.

\begin{figure}
\includegraphics[width=8.5cm]{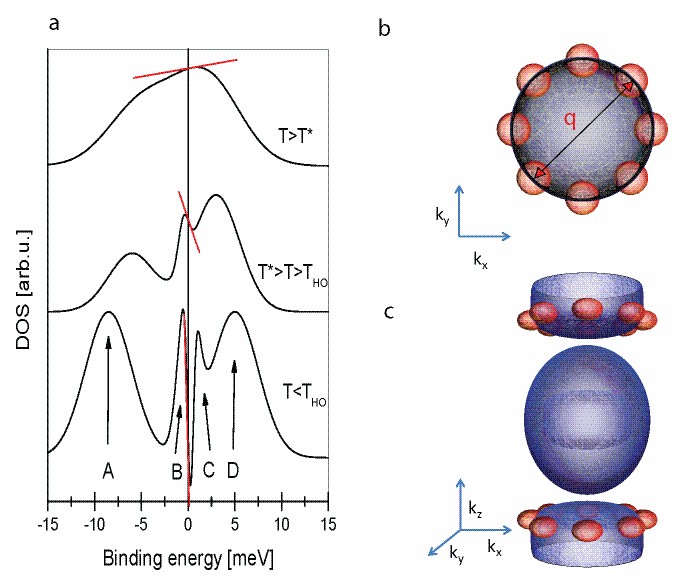}
\caption{
(color online) Model of the Fermi surface renormalization in {\ur}.
{ \bf {a,}} evolution of the HO gap inside the hybridization gap. Red line corresponds to the slope of density of states extracted from thermoelectric power measurements \cite{Hundley}, a very sensitive probe of the integrated density of states (DOS) at the
Fermi level. Dip in DOS between B and C on the scale of few meV is smaller than the resolution of our tr-ARPES instrument and hence we rely on transport measurements at these energies, and on time-resolved data for deduction of the gap structure. {\bf{b,}} in-plane cut through the Fermi surface of {\ur} illustrating the location in k-space of the two QP spots measured here, as indicated by vector q. Both these spots and the ones along (1 0 0) direction are independently found by STM \cite{Schmidt}. {\bf{c,}} the side view of the proposed Fermi surface in HO state, modified after \cite{Oppeneer2010}; the exact size of the pockets along k$_z$ direction remains unknown. }
\end{figure}

Measurements performed with tr-ARPES at 12 K and at 19 K, below and above HO transition, show the well-resolved QP structure, which exists only in the HO phase and a less pronounced QP structure existing above T$_0$  (see Fig. 2 a and b). Location of the  $q_{<110>}$ vector is shown in panel c.  Here the density of states is only slightly elevated at 19 K, and significantly elevated at 12 K. ARPES was performed in order to establish the occupied side of the electronic band structure with high resolution (Fig. 2 panel d) and indicate the existence of a narrow heavy band above the Fermi level. The lifetime measurement indicated in Fig. 2 and presented in Fig. 3 shows that the QP lifetime at 19 K is almost an order of magnitude shorter than at 12 K, indicating much broader peaks above T$_0$ than in the HO state.  The possible enhancement of QP lifetime at T$_0$ was derived from transport measurements \cite{Behnia}, but with unknown magnitude.  We propose that the 19 K structure corresponds to the hybridization gap, which is different from HO gap. To summarize the differences, first, the hybridization gap opens around the coherence temperature T*, which is at least two times larger than T$_0$ \cite{Palstra,Schlabitz,Maple}. Next, the estimated size of the hybridization gap is roughly two times larger than HO gap. Finally, the QPs contributing to the edges of the hybridization gap are short-lived and hence much broader than the HO gap structure.

The model of Fermi surface renormalization at T$_0$ can now be proposed as follows (Fig.4). At high temperature T$>$T*, the metallic system shows full density of states around the Fermi level.
Below T* but above T$_0$, T*$>$T$>$T$_0$, the hybridization gap opens in the density of states \cite{Oppeneer2010}, with structure determined by two broad peaks, A and D, and some density of states at the Fermi level. The size of the hybridization gap between 10 meV and 20 meV is
estimated from optical reflectivity and universal scaling\cite{BauerPRL}.  At T$_0$ the HO gap opens, removing large amount of the density of states from the Fermi level and significantly reducing entropy. The size of this gap determined by peaks B and C is between 5 meV and 10 meV\cite{Wiebe,Santander,Schmidt,Aynajian}. Peaks C and D, essential in establishing the structure of the Fermi surface renormalization, are evidenced by the QP lifetime change across the transition derived from the time-resolved and momentum-resolved data taken at 12K and 19K, respectively. The momentum-resolved cut through the Fermi surface is shown in Fig. 4 b and a full 3D renormalization can be seen in Fig. 4c.

We find that within the well-defined regions in momentum space the QP lifetime increases dramatically below T$_0$ in an anomalous manner not seen in other heavy fermion materials. Partial gapping of the QPs near the Fermi surface leads to suppressed recombination channels for the excited QPs. The structure of the QPs in momentum space strongly hints at momentum-dependent interactions being responsible for the HO gapping. We conclude that the HO state is caused by a rapid renormalization of the FS through gapping at specific spots. Our novel tr-ARPES allows a direct, femtosecond-scale visualization of the formation of the long-lived QPs as the driver of HO. The HO gap occurs as a momentum-dependent modification of a momentum-independent hybridization gap, and it constitutes the order parameter for the HO transition. These findings are not inconsistent with the itinerant character of the hidden order proposed by commensurate or incommensurate models \cite{Chandra, Elgazzar, Wiebe, Hassinger, Schmidt, Aynajian, Tripathi,Balatsky, Dubi, Oppeneer2010}, and they underscore the need for new, advanced many-body modelling, taking into account the QP physics in the context of both T* and T$_0$ energy scales. 
 
We also sketch a model based on time-resolved ARPES data and transport measurements. Our model allows the reconciliation of apparently 
contradictory previous photoemission results, by acknowledging that the 21.2eV \cite{Santander}, 7eV \cite{Yoshida} and our 34eV photon energy datasets represent sampling of different parts of the 3D Fermi surface, and with very different 5f photoionization cross sections, ranging from 0 barns at 7eV to few million barns at 34eV \cite{adinf}. Our observations provide an explanation for the entropy removal at  T$_0$, and point towards the symmetry breaking through momentum-dependent scattering at $q_{<110>}  =  0.56 \pm  0.08$ {\an} as the source of second-order transition. 

These results place further constraints on the possible theories of HO and will aid and stimulate the long-standing quest to uncover the exact nature 
of the enigmatic HO. The ability to visualize the evolution of the QP lifetime in energy-momentum space opens new and exciting possibilities for heavy-fermion and beyond research by investigating the complex time-dependent near-Fermi-level electronic structure via femtosecond dynamics.  

We thank A.F. Santander-Syro and G. Lander for helpful discussions. Work at LANL was performed under the 
auspices of the US DOE, LANL LDRD and UCOP-TR01 Programs. Research was partially conducted at the Synchrotron Radiation Center, supported by the NSF under Award No. DMR-0537588. P.M.O. was supported
through the Swedish Reserach Council (VR) and
EU-JRC ITU. P.R. was supported by the US DOE BES award DEFG02-84ER45872.

* corresponding author: tomasz@lanl.gov

{}

\end{document}